# Defect thermodynamics and kinetics in thin strained ferroelectric films: the interplay of possible mechanisms

Anna N. Morozovska[1,2], Eugene A. Eliseev[2], P.S.Sankara Rama Krishnan[3], Alexander Tselev[4], Evgheny Strelkov[4], Albina Borisevich[4], Olexander V. Varenyk[5], Nicola V. Morozovsky[1], Paul Munroe[3], Sergei V. Kalinin[4*] and Valanoor Nagarajan[3†]

[1]Institute of Physics, [2]Institute of Problems for Material Sciences, NAS of Ukraine, 03028 Kiev, Ukraine

[3]School of Materials Science and Engineering, the University of New South Wales, Sydney, NSW2052 Australia.

[4]The Center for Nanophase Materials Sciences, Oak Ridge National Laboratory, Oak Ridge, TN 37922

[5]Taras Shevchenko Kiev National University, Radiophysical Faculty, 4g, pr. Akademika Hlushkova, 03022 Kiev, Ukraine

## Abstract

We present a theoretical description of the influence of misfit strain on mobile defects dynamics in thin strained ferroelectric films. Self-consistent solutions obtained by coupling the Poisson's equation for electric potential with continuity equations for mobile donor and electron concentrations and time-dependent Landau-Ginzburg-Devonshire equations reveal that the Vegard mechanism (chemical pressure) leads to the redistribution of both charged and electro-neutral defects in order to decrease the effective stress in the film. Internal electric fields, both built-in and depolarization ones, lead to a strong accumulation of screening space charges (charged defects and electrons) near the film interfaces. Importantly, the corresponding screening length is governed by the misfit strain and Vegard coefficient. Mobile defects dynamics, kinetics of polarization and electric current reversal are defined by the complex interplay between the donor, electron and phonon relaxation times, misfit strain, finite size effect and Vegard stresses.

[*] Corresponding author

[†] Corresponding author

## I. Introduction

The thermodynamic stability of point defects, such as vacancies and the resulting cation kinetics, remains a highly intriguing and controversial area in functional oxide thin films and interfaces . [1, 2, 3] Interface regions naturally provide a rich tapestry of physical and chemical defect phenomena, such as those related with misfit dislocations, vacancies and other lattice defects. [4, 5, 6, 7] In particular epitaxial ferroic metal oxide thin films with the perovskite $ABO_3$ structure (bismuth ferrite-$BiFeO_3$, lead zirconate titanate - $(Pb,Zr)TiO_3$, strontium titianate – $SrTiO_3$ etc) are extremely sensitive to the presence of mobile oxygen vacancies and cations. It is well known that such entities have a very strong and non-trivial effect on their electro-physical, polar and magnetic properties. [8, 9]

Over the years, numerous models based on factors such as misfit strain and kinetics of ion mobility have been developed to analyse and understand various factors governing the interfacial chemical stability (see e.g. Ref. [[10], [11]]). For example, Stephenson and Highland [12] developed a thermodynamic theory of the ferroelectric phase transition in an ultrathin film in equilibrium with a chemical environment that supplies ions to screen bond surface charges caused by the spontaneous polarization. Similarly, Shenoy et al. [13, 14] considered the impact of charged vacancies and cations on the formation of space charge layers, which produce strains that substantially alter thermodynamic equilibrium near surfaces in ionic solids. The thermodynamic approach was then extended to ferroelectric films with different doping levels by donors. [15] While the previous reports account for the epitaxial misfit strain in the Landau expansion on polarization powers, how the misfit strain affects oxygen vacancies redistribution has not yet been explained. On the other hand the kinetics of vacancy diffusion across interfaces (without the influence of misfit strain) in other materials systems such as multi-component alloys and electrode-electrolyte interfaces has been modelled.[1, 16, 17]

With the continuing push for oxide nanoelectronic devices with interface-driven functionalities the interplay between defects and misfit strain can no longer be ignored. For example, Kalinin and Spaldin recently offered a perspective on how tuning the defect concentration and profiles could in itself lead to new functionalities in metal oxide materials systems.[18] First principle computations on calcium manganite recently demonstrated that epitaxial misfit strain can be exploited as a route to engineering vacancy ordering in epitaxial thin films.[19] Most critically, it was shown that point defect formation is also a likely strain-relaxation mechanism.

Previously, we reported about misfit strain driven La cation inter-diffusion across the epitaxial $BiFeO_3$ (BFO) thin film from the interface with $LaAlO_3$ (LAO).[20] We observed cation intermixing at the interface using an aberration corrected scanning transmission electron microscopy-electron energy loss spectroscopy (STEM – EELS). The experimentally observed cation intermixing over a length scale of ~ 2 nm was explained based on the driving forces generated under the influence of the misfit strain. A Ginzburg-Landau-Devonshire (LGD) thermodynamic model combined with Sheldon and Shenoy formalism [13, 14, 15] was used to explain the cation intermixing.

Generally speaking, the influence of mechanical and electrical boundary conditions at an oxide interface on defect migration still remains a theoretical challenge. The local redistribution, $\delta N$, of mobile ions concentration, $N$, caused by electromigration (electric field-driven) and diffusion (concentration gradient-driven) mechanisms changes the lattice molar volume. The changes in volume result in local electrochemical strains, $\delta u \propto W \delta N$, which can be regarded as "Vegard strains" and $W$ is a Vegard coefficient. [21] The Vegard mechanism (chemical pressure) plays a decisive role in the origin of local strains caused by the point defects kinetics in solids.[22, 23, 24, 25, 26] Analytical models [27, 28, 29, 30, 31] proved that the coupling between ionic redistribution and Vegard strains can give rise to the local strain response in mixed electronic-ionic Li-containing conductors and $SrTiO_3$ with charged defects. In all of the abovementioned Refs [1-12**Ошибка! Закладка не определена.**, 27-31] the critical influence of the misfit strain on the oxygen vacancies and their redistribution was not considered. The simple theoretical model used for the description of experimental results [20] considered only charged vacancies thermodynamics. Naturally, we need to study additional variables (hysteresis loops, etc) to obtain a fuller picture of the correlation between the functional properties and interfacial thermodynamics and kinetics of both, charged and neutral species in thin ferroelectric films. This motivated us to perform the present theoretical study, where the main aim was to explore how the misfit strain can influence the gradient of defect concentration and consequently device properties, such as exploiting the polarization reversal and electric current kinetics.

Our theoretical approach is based on the solution of the Poisson equation for electric potential coupled with kinetic equations for donor and electron concentrations and time-dependent Landau-Ginzburg-Devonshire (LGD) equation for the ferroelectric polarization. Vegard stresses and misfit strain lead to the shift of the donor and electron chemical potentials in kinetic equations and dielectric stiffness in LGD equation. To illustrate the approach we chose BFO film clamped on a rigid oxide substrate, since the system is one of the most promising for advance applications in nanoelectronics. [32, 33, 34]

The original part of the paper is organized as follows. The problem and basic equations are stated in section 2. Analytical descriptions of the system behaviour in thermodynamic equilibrium are evolved in section 3. Ferroelectric polarization reversal and electric current kinetics are analysed in section 4. Section 5 summarizes briefly our results. Evident forms of the LGD thermodynamic potential, coupled equations for ferroelectric polarization reversal and electric current kinetics in dimensionless variables, and calculation details of the equilibrium concentration of neutral defects are listed in the **Supplementary Materials.**

## II. Problem statement and basic equations

Let us consider an electroded oxide ferroelectric thin film without a domain structure clamped by a rigid substrate (see **Figure 1**).

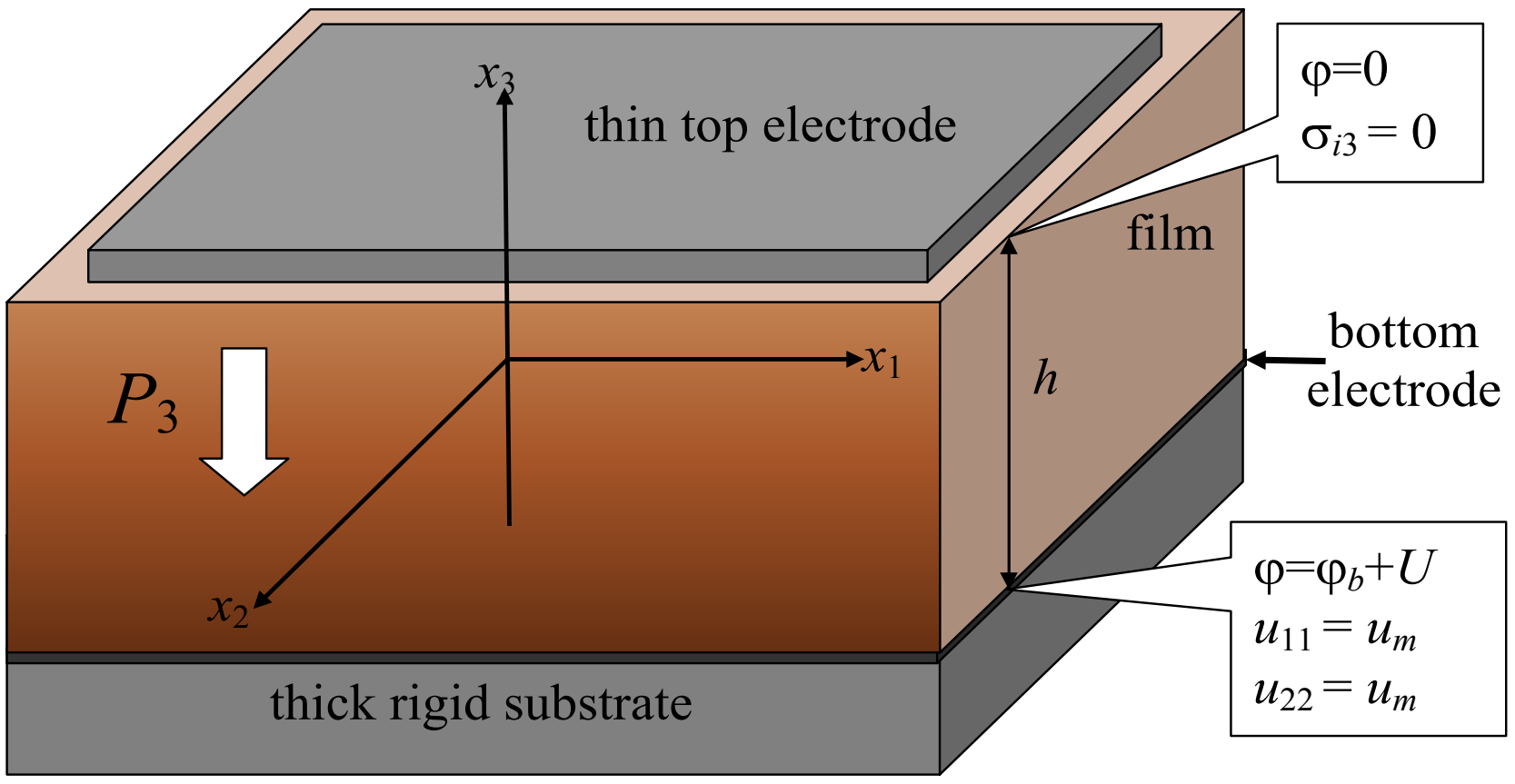

**Figure 1.** Schematics of a ferroelectric epitaxial film clamped on a rigid substrate. Note that a thermodynamics of a similar system without thin top electrode was considered in Ref.[20].

In the case all physical quantities depend only on the distance $x_3$ from the film-substrate interface (1D problem). Epitaxial misfit strain $u_m$ can exist at the film/substrate interface. Mobile positively charged point defects (e.g. oxygen vacancies or cations) and free electrons are regarded inherent to the film. The ferroelectric polarization $P_3(x_3)$ and mobile charge carriers redistribution can create the internal electric field in the film, $E_3 = -\partial\varphi/\partial x_3$ ,

where φ is the corresponding electric potential. Potential φ is described by the Poisson equation in a self-consistent way:

$$\varepsilon_0\varepsilon_b\frac{\partial^2\varphi}{\partial x_3^2}=\frac{\partial P_3}{\partial x_3}-e\left(\sum_k Z_k N_k^+(\varphi)-n(\varphi)\right) \tag{1a}$$

Here $\varepsilon_0$=8.85×10$^{-12}$ F/m, the dielectric permittivity of vacuum; $\varepsilon_b$ is a background permittivity of ferroelectric,[35] $P_3$ is a ferroelectric polarization, electron density is $n$, ionized defect concentration is $N_k^+$, $e$=1.6×10$^{-19}$ C the electron charge, $Z_k$ is the defect charge (that is equal to zero for uncharged vacancies or isovalent impurities). For the sake of clarity let us suppose that electric potential satisfies the fixed boundary conditions at the electrodes

$$\varphi(0)=U(t)+\varphi_b, \qquad \varphi(h)=0, \tag{1b}$$

which corresponds to the electroded film of thickness $h$ with perfectly conducting thin top electrode. The (constant or periodic) voltage $U$ is applied to the bottom electrode, $\varphi_b$ is built-in potential (if any originated from e.g. Shottky contact between the film and bottom electrode).

The donor dynamics are described by the corresponding continuity equation supplemented with boundary conditions. Continuity equation for ionized donor concentration $N_k^+$ is [31]:

$$\frac{\partial N_k^+}{\partial t}+\frac{1}{eZ_k}\frac{\partial J_k^d}{\partial x_3}=0, \tag{2}$$

where the current $J_k^d$ is proportional to the gradients of the carrier electrochemical potentials levels $\zeta_k$ as $J_k^d=-eZ_k\eta_k N_k^+\left(\partial\zeta_k/\partial x_3\right)$, where $\eta_k$ is the ions/vacancies mobility coefficient that is regarded constant. The electrochemical potential level $\zeta_k$ in expression for donor current is defined as:

$$\zeta_k\approx-E_{dk}-W_{ij}^k\sigma_{ij}+eZ_k\varphi+k_BT\ln\left(\frac{N_k^+}{N_k^0-N_k^+}\right). \tag{3}$$

Where $E_{dk}$ is the $k$-type donor level, misfit-dependent elastic stress tensor is $\sigma_{ij}$, $T$ is the absolute temperature, $k_B$ is a Boltzmann constant, $W_{ij}^k$ is the Vegard strain tensor of (other equivalent names are chemical pressure [32] and elastic dipole [36] tensor) hereinafter regarded as diagonal [37, 38] and $W_{ij}^k=W^k\delta_{ij}$ ($\delta_{ij}$ is delta Kroneker symbol). The absolute value of $W$ for $ABO_3$ compounds was estimated as $|W|$ = (1 ÷ 10) Å$^3$ following refs. [36]. The maximal possible concentration of donors ($N_k^0$), takes into account steric effects .[39, 40] For numerical estimates one should assume that $N_k^0 \equiv a^{-3}$, where $a^3$ is the maximal volume allowed per donor centre. From Equation (3), donor concentration is $N_d^+ = N_k^0\left(1 - f\left(E_{dk} + W_{ij}^k\sigma_{ij} - eZ_k\varphi + \zeta_d\right)\right)$, and thus $N_d^+ < N_k^0$ because the Fermi-Dirac distribution function $f(x) = \left(1 + \exp(x/k_B T)\right)^{-1}$ is always smaller than unity at finite temperatures. Electrons are regarded to be infinitely small and the strains associated with electron crowding are not considered here, i.e. the deformation potential effect is neglected.

After establishing the methodology for estimating the local electrochemical potential based on donor concentration, we turn attention towards fixing the boundary conditions. The boundary conditions for the ionic/vacancies current at the film-substrate interface $x_3 = 0$ is taken in the linearized Chang-Jaffe (CJ) form, $\left.\left(J_k^d - v_k\left(N_k^+ - N_k^S\right)\right)\right|_{x_3=0} = 0$, $v_k$ is a positive rate constant related with the surface recombination velocity. [41] The CJ condition contains the continuous transition from the "open" interface ($v_k \to \infty \Rightarrow N_k^+ = N_k^S$) to the interface limited kinetics ($0 < v_k < \infty$) and "completely blocking" interface ($v_k = 0$). The surface $x_3 = h$ is regarding blocking for mobile donors and so $J_k^d(h) = 0$.

To calculate electron dynamics, one should solve the corresponding kinetic equation supplemented with boundary conditions. Under negligibly small impact of the electron hopping and recombination-generation process, continuity equation for electrons is:

$$\frac{\partial n}{\partial t} - \frac{1}{e}\frac{\partial J_3^e}{\partial x_3} = 0 \qquad (4)$$

Where the electron current $J_e = e\eta_e n\left(\partial\zeta_e/\partial x_3\right)$, $\eta_e$ is the electron mobility coefficient. In continuous approximation for the concentration of the electrons in the conduction band the electrochemical potential level $\zeta_e$ in expression for electron current is defined as:

$$\zeta_e \approx E_C + k_B T\, F_{1/2}^{-1}\left(\frac{n(\varphi)}{N_C}\right) - e\varphi, \qquad (5)$$

where the electro-chemical potential $\zeta_e$ tends to the Fermi energy level $E_F$ in equilibrium, $E_C$ is the bottom of conductive band, $F_{1/2}^{-1}$ is the function inverse to the Fermi integral $F_{1/2}(\xi) = \frac{2}{\sqrt{\pi}}\int_0^\infty \frac{\sqrt{\zeta}d\zeta}{1 + \exp(\zeta - \xi)}$; effective density of electron states in the conductive band in the effective mass approximation is $N_C = \left(\frac{m_n k_B T}{2\pi\hbar^2}\right)^{3/2}$, where an electron effective mass is $m_n$.[42] Electron density can be calculated from Eq.(5) as $n = N_C F_{1/2}\left((e\varphi + \zeta_e - E_C)/k_B T\right)$.

The CJ boundary conditions for the electron current are $J_e(0) = v_0\left(n(0,t) - n_0\right)$ and $J_e(h) = -v_1\left(n(h,t) - n_1\right)$. Numerical values of $v_{0,1}$ are determined by the electrode and film material. If the rate constants are infinitely high, then the equilibrium electron density at the contacts is fixed by the electrodes and independent of the applied voltage.[43, 44]

In order to define the electrochemical potential in expression (3) and thus to determine the donor distribution, one should find the elastic stress tensor in the film allowing for the relevant boundary conditions. Mechanical boundary conditions corresponding to the epitaxial thin film clamped on a rigid cubic substrate are conventional, $\sigma_{3i}(0) = 0$ at the top free surface $x_3 = 0$, and $u_{11}(h) = u_{22}(h) = u_m$ at the film-substrate interface $x_3 = h$, $u_m$ is the film-substrate misfit strain. In the case where dislocations are absent, the stress field has the form:

$$\sigma_{13} = \sigma_{23} = \sigma_{33} = 0, \qquad \sigma \equiv \sigma_{11} + \sigma_{22} = \sigma_0 - \frac{2W^k\left(N_k^+ - N_{ke}^+\right)}{s_{11} + s_{12}}. \qquad (6)$$

The effective bi-axial stress $\sigma_0$ originates from misfit, polar long-range ordering and flexoelectric coupling, $N_{ke}^+$ is the equilibrium constant concentration of the charged defects (see [45, 46] and **Appendix A of Suppl. Mat**). It can be stated that the impact of the flexoelectric coupling on the shear stress can be ignored only for the case of purely out-of-plane polarization $\mathbf{P} \equiv P_3(x_3)$. Exactly the case is considered below and for this case the stress is:

$$\sigma_0(x_3) = \frac{2u_m - 2Q_{12}P_3^2(x_3)}{s_{11} + s_{12}} \qquad (7)$$

Here $Q_{ijkl}$ is the electrostriction tensor coefficient, $s_{ij}$ are elastic compliances. For typical perovskite-type ferroelectrics with $(Q_{11} + Q_{22}) > 0$ and $Q_{12} < 0$, out-of-plane polarization can be stabilized by compressive strains $u_m < 0$.[47] and therefore compressive strains will be considered below.

Note, that the stress gradient in ferroelectrics-antiferrodistortive perovskites (like BFO) can be caused by the octahedral tilt gradient across the interface via rotostriction mechanism as estimated in the **Appendix A of Suppl. Mat.** More importantly, since the stress is misfit-dependent, electrochemical potential level (3) and donor concentration become misfit-dependent in a self-consistent manner. The dependence of the donor concentration on the misfit-strain can lead to their misfit-driven diffusion.

To calculate the stress from Eq.(7) one requires to know the polarization distribution $P_3(x_3)$. Inhomogeneous spatial distribution of the polarization $P_3(x_3)$ can be determined from the time-dependent LGD equation:

$$\Gamma\frac{\partial}{\partial t}P_3+\alpha_R(T,x_3)P_3+a_{11}P_3^3-g_{11}\frac{\partial^2 P_3}{\partial x_3^2}=-\frac{\partial\varphi}{\partial x_3} \tag{8}$$

Where $\Gamma$ is the Khalatnikov coefficient determined by the phonon relaxation time, $\alpha_R(T,x_3)=a_1(T)-2Q_{33ij}\sigma_{ij}(x_3)$, and $a_i$ and $a_{ij}$ are the coefficients of the LGD potential expansion on the polarization powers. Corresponding boundary conditions are, $\left.\left(P_3-\lambda_P\frac{\partial}{\partial x_3}P_3\right)\right|_{x_3=0}=0$ and $\left.\left(P_3+\lambda_P\frac{\partial}{\partial x_3}P_3\right)\right|_{x_3=h}=0$. The geometrical sense of extrapolation length and $\lambda_P$ is described in Ref.[48]. The lengths are determined by the surface energy that depends on the surface state and is poorly known for ferroelectrics. The physically realistic range is 0.5 – 2 nm.[49] The boundary conditions reflect the surface energy contribution into the polarization vector components slope near the surface.

Thus, in the proposed model, the mathematical statements given by Eqs.(1)-(8), are self-consistent because it allows one to calculate the influence of the misfit strain on the elastic stress distribution (via Eq.(7)), then the stress distribution on the donor redistribution (via Eqs.(1)-(3)), and finally the donor redistribution influence on the electric potential (via Eq.(1)), and polarization (via Eq.(8)) in a self-consistent manner. Using the formulation of the basic framework given by Eqs.(1)-(8) the influence of the misfit strain on the cation or vacancies interdiffusion in heteroepitaxial oxide thin films can be established comprehensively.

### III. The system behaviour in thermodynamic equilibrium

In order to shed more light on the film state at very slow variations of applied voltage, next we consider the system in thermodynamic equilibrium. Examination of formulation (1)-(8) for the equilibrium case gives us additional grounds for its applicability in the kinetic case.

So, let us analyse the system behaviour in the thermodynamic equilibrium at zero external voltage, $U=0$. Below we calculate the concentration of ionized donors, electrostatic potential, elastic stress and polarization distribution across the film depth z. It is necessary to know these physical quantities behaviour for the characterization of electrophysical properties such as origin of the space charge layers, static conductivity, I-V characteristics, polarization hysteresis loops shape, corresponding remanent polarization and coercive field in a strained ferroelectric thin film.

Following the Shenoy et al. formalism [13-15], the equilibrium defect concentration consistent with the statistics of chemically diluted solutions can be derived after the solution of Eq.(3) with respect to $N_d^+$, allowing for the fact that the position of donors' electrochemical potential $\zeta_d$ in Eq.(3) is given by the Fermi energy, e.g. $\zeta_d = -E_F$ in thermodynamic equilibrium. Thus we obtained:

$$N_d^+(\sigma_{ij},\varphi) = N_d^0\left(1 - f\left(E_d + W_{ij}\sigma_{ij} - eZ_d\varphi - E_F\right)\right) \tag{9a}$$

Fermi-Dirac distribution function is $f(x) = \left(1 + \exp(x/k_BT)\right)^{-1}$. Let us recall, that $N_d^0 \propto \left(10^{24} - 10^{26}\right)$ m$^{-3}$ is the maximal steric concentration of defect atoms.

Note, that Equation (9a) is valid beyond Boltzmann approximation and thus accounts for the steric effects. In Boltzmann approximation Eq.(9a) gives concentration variation $N_d^+ = N_{de}^+\exp\left(\left(W_{ij}\sigma_{ij} - eZ_d\varphi\right)/k_BT\right)$, where $N_{de}^+ = N_d^0\exp\left(\left(E_d - E_F\right)/k_BT\right)$ is the equilibrium concentration of ionized donors. The concentration of the electrons in the thermodynamic equilibrium can be derived directly from Eq.(5) under the condition $\zeta_e = E_F$. By solving the Eq.(5) with respect to $n$ we obtained the dependence of the concentration on the electric potential as:

$$n = N_C F_{1/2}\left(\left(e\varphi + E_F - E_C\right)/k_BT\right). \tag{9b}$$

When the dependence of donors and electrons concentration on electrostatic potential are determined by Eqs.(9a and 9b), one can find numerically the electrostatic potential from Eq.(1a) with boundary conditions (1b) in a self-consistent manner.

Approximate analytical results can be obtained from the Debye approximation that is valid under the condition $\left|eZ_d\varphi\right| < k_BT$. In the approximation $N_d^+ \approx N_{de}^+\exp\left(W_{ij}\sigma_{ij}/k_BT\right)\left(1 - eZ_d\varphi/k_BT\right)$, $n \approx n_0\left(1 + e\varphi/k_BT\right)$ and electroneutrality at zero potential gives the condition $n_0 = N_{de}^+\exp\left(W_{ij}\sigma_{ij}/k_BT\right)$. In the Debye approximation Eq.(1a) becomes $\frac{\partial^2\varphi}{\partial x_3^2} - \frac{\varphi}{R_d^2} = \frac{1}{\varepsilon_0\varepsilon_b}\frac{\partial P_3}{\partial x_3}$, and its solution is:

$$\varphi(x_3)=\varphi_b\frac{\sinh((h-x_3)/R_d)}{\sinh(h/R_d)}-\frac{R_d}{\varepsilon_0\varepsilon_b}\left(\int_0^{x_3}\left(\frac{dP_3}{d\xi}\right)G(\xi,z)d\xi+\int_{x_3}^{h}\left(\frac{dP_3}{d\xi}\right)G(z,\xi)d\xi\right). \qquad (10)$$

The Green's function is $G(\xi,z)=\frac{\sinh(\xi/R_d)\sinh((h-z)/R_d)}{\sinh(h/R_d)}$. The first term originated from the built-in potential, the second one originated from the depolarization field. In Equation (10) the Debye screening radius $R_d$ is introduced as:

$$R_d=R_d^0\exp\left(-\frac{W\overline{\sigma}_0}{2k_BT}\right), \qquad (11)$$

The typical values of radius $R_d^0=\sqrt{\frac{\varepsilon_0\varepsilon_bk_BT}{e^2(Z_d+1)N_{de}^+}}$ is $(1-10)$nm for the donor charge $Z_d=1$, T=300 K, background permittivity $\varepsilon_b\propto 5$ and defect equilibrium concentration $N_{de}^+=(10^{22}-10^{24})$ m$^{-3}$. Vegard strain coefficient absolute value can be taken as $|W|\propto$ (1 – 10) Å$^3$. Designation $\overline{\sigma}_0$ in Eq.(11) is the effective stress. The average stress can be obtained by the averaging of Eq.(7) over the film thickness *h*:

$$\overline{\sigma}_0=2\frac{u_m-Q_{12}\overline{P_3^2}}{s_{11}+s_{12}}. \qquad (12)$$

Since the expressions for effective stress (12) and potential (10) are polarization-dependent one should determine the polarization distribution from e.g. Eq.(8) to make the derived expressions self-consistent. In the static case the solution of Eq.(8) for polarization is [50]:

$$P_3(x_3)=\overline{P}_3\left(1-\frac{\cosh(x_3/L_P)}{\cosh(h/2L_P)+(\lambda_P/L_P)\sinh(h/2L_P)}\right). \qquad (13)$$

Solution (13) near the interface $x_3=0$ acquires the following form $P_3(x_3)\approx\overline{P}_3\left(1-\frac{\exp(-\sqrt{2}x_3/L_P)}{1+\sqrt{2}\lambda_P/L_P}\right)$. Here $\overline{P}_3$ is the spontaneous values of the order parameter in the film, which are misfit-dependent in accordance with LGD-theory. Correlation lengths are introduced as $L_P(T)$. Note, that the polarization component normal to the interface, $P_3(x_3)$, is suppressed by the depolarization field, and thus its gradient typically extend only for 0.1 – 0.5 nm. The corresponding correlation length $L_P(T)$ is about, or smaller than, the lattice constant. For this case only the strain gradient gives rise to the diffusion of both charged and uncharged species. Note that in fact the LGD-expansion coefficient is renormalized by misfit strain and Vegard effect in Eq.(8), as $\alpha(T)=\alpha_T\left(T-T_C^b\right)-\frac{4Q_{12}\left(u_m+W\left(N_d^+-N_{de}^+\right)\right)}{\left(s_{11}+s_{12}\right)}$. This fact allows us to estimate the dependence of the average remanent polarization $\overline{P_3}\left(E_3^{ext}=0\right)$ and coercive field $E_{cr}^T$ on misfit strain $u_m$ and Vegard coupling $W$ as:

$$\overline{P}_3 \approx P_S^T\sqrt{\left(1-\frac{T}{T_C^b}+\frac{4Q_{12}\left(u_m+W\left(\overline{N}_d^+-N_{de}^+\right)\right)}{T_C^b\alpha_T\left(s_{11}+s_{12}\right)}-\frac{h_{cr}}{h}\right)}, \qquad (14a)$$

$$E_{cr} \approx E_{cr}^T\left(1-\frac{T}{T_C^b}+\frac{4Q_{12}\left(u_m+W\left(\overline{N}_d^+-N_{de}^+\right)\right)}{T_C^b\alpha_T\left(s_{11}+s_{12}\right)}-\frac{h_{cr}}{h}\right)^{3/2}. \qquad (14b)$$

Where $P_S^T$ are $E_{cr}^T$ the spontaneous polarization and thermodynamic coercive field of bulk ferroelectric at temperature 0 K, $h_{cr}=\frac{L_P}{\varepsilon_0\varepsilon_b\alpha_T T_C^b\left(L_P+\lambda_P\right)}$ is the critical thickness of the finite size induced phase transition into the paraelectric phase.

When the evident expression for electric potential (Eq.(10)) effective stress (Eq.(12)) and average polarization (Eq.(14a)) are known one can calculate the concentration of ionized donors and electrons from Eqs.(9). Here we should note that Eq.(10) as Debye approximation that has satisfactory accuracy only under the condition $\left|eZ_d\varphi\right|<k_BT$. Boltzmann approximation typically works at $\left|eZ_d\varphi\right|<10k_BT$; with further increase of $\left|eZ_d\varphi\right|$ steric effects included in Eq.(9a) start. Expressions (10)-(14) valid in the Debye approximation can explain the system properties on the analytical level. Numerical simulations (results are presented below) proved that the strong dependence on $u_m$ remained beyond the limits of Debye or Boltzmann approximation applicability, namely in the case $\left|eZ_d\varphi\right| >> k_BT$.

After establishing the mathematical derivatives of the ferroelectric system in thermodynamic equilibrium, the dependence of the electric potential, electrons, ionized defects and polarization distribution on the misfit strain and Vegard tensor should be studied. This can be established by understanding the dependence of $u_m$ and $W$ on all these properties. The dependencies illustrated by **Figures 2-4** were calculated numerically from Eqs.(1)-(9) without any suggestions of the Debye or Boltzmann approximation validity. In order to study the net effect of depolarization field we assume zero built-in potential ($\varphi_b = 0$) during the calculations.

Depolarization field potential $\varphi$, polarization $P_3/P_S$, relative concentration of ionized donors $N_d^+/N_{de}^+$ and electrons $n/N_{de}^+$ profiles calculated for a BFO ferroelectric film under different compressive misfit strains $u_m$ are shown in **Figure 2.** The position 0 nm refers the substrate-film interface, while 20 nm refers the top end of the film. Plots **a** and **b** show the distribution of $\varphi$ and $P_3/P_S$ across the film thickness, which are almost indifferent to Vegard coefficient values in the range {−1, +1} Å$^3$ and strongly asymmetric with respect to the film centre. Actually, different behaviour of the potential and polarization near the film surfaces is related to the different width and shape of electron and donor screening layers (shown in the plots **c** and **d**). Let us underline that the length of the polarization gradient near two film surfaces is strongly different, it is about 1 nm near the surface $x_3 = 0$ enriched by ionized donors, and about 5 nm near $x_3 = h$ enriched by electrons. Consequently, the actual scale is determined by the width of the screening space charge layer, not by the correlation length $L_P \approx 0.1 - 0.5$ nm or equal extrapolation lengths $\lambda_P = 0.1 L_P$. Plots **c** and **d** show corresponding distributions of the screening charge concentration, $n/N_{de}^+$ (solid curves) and $N_d^+/N_{de}^+$ (dashed curves) for the positive and negative Vegard coefficient respectively. A steric effect leading to the plateau on the ionized donor distribution near the electrode $x_3 = 0$ is pronounced irrespective of the misfit value. The widths of the electron and donor screening layers depend on misfit strain. Polarization maximal values increase under the increase of compressive misfit strain in qualitative agreement with Eq.(14a).

Importantly, the theoretical results illustrated by **Figure 2** can explain the experimental observations.[20] In particular, one can see that electric potential and donor concentration gradient noticeably depends on misfit value. For a BFO thin film epitaxially clamped on a rigid LAO substrate studied in experimentally [20] the misfit was rather high ($u_m = +4.5$%), La ions were donors, which migrate into the film from a substrate. For this case our modelling predicts that the donor concentration decreases 10 times at distances about 5-10 nm from the BFO/LAO interface. This in agreement with experimental observations [20].

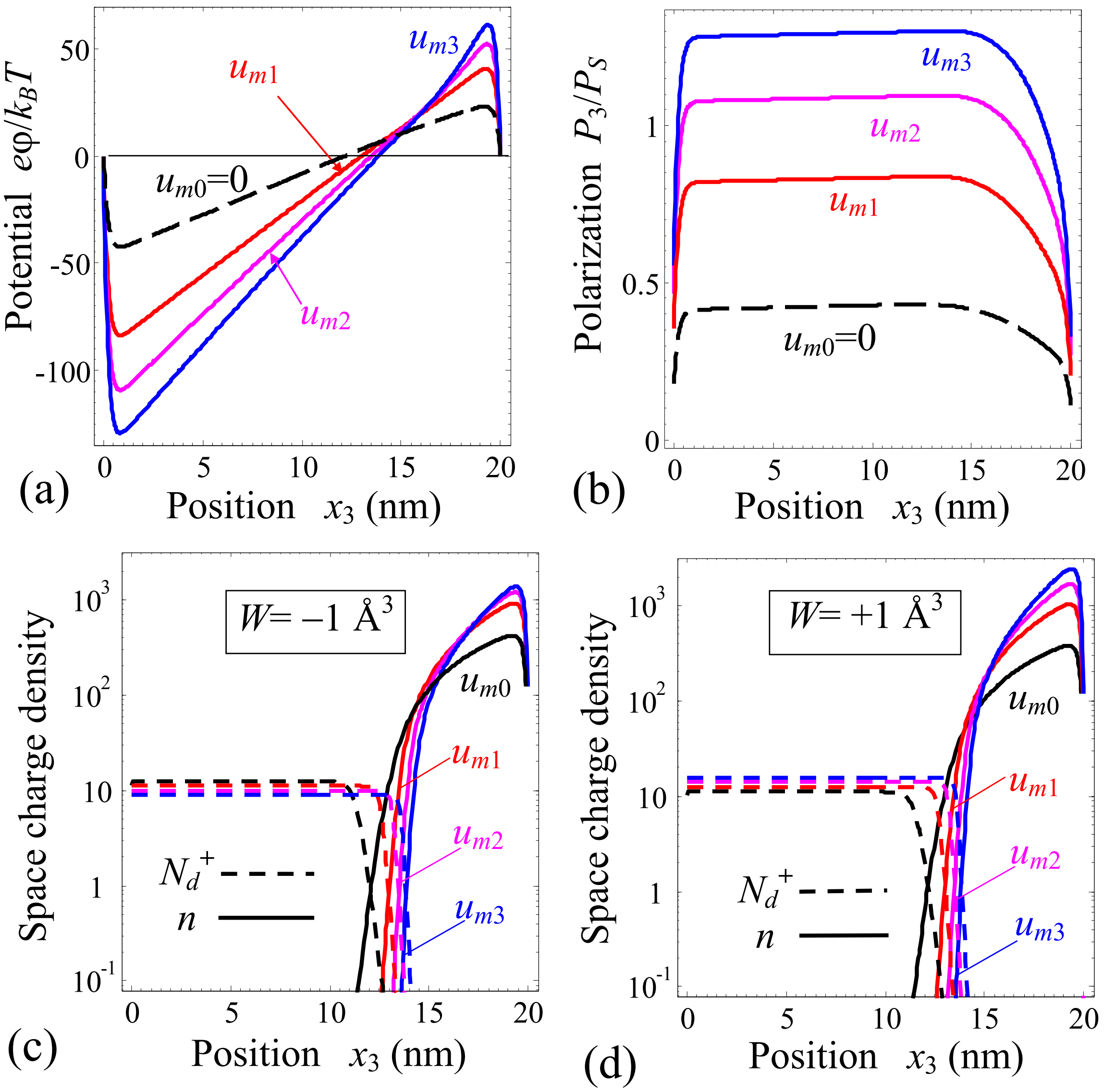

**Figure 2.** Potential $\varphi$ **(a),** polarization $P_3/P_S$ **(b)**, relative concentration of ionized donors $N_d^+/N_{de}^+$ and electrons $n/N_{de}^+$ **(c,d)** profiles calculated for BFO film of thickness $h$= 20 nm, absence of the built-in potential ($\varphi_b = 0$), different misfit strains $u_{m0} = 0$, $u_{m1} = -1\%$, $u_{m2} = -2\%$, $u_{m3} = -3\%$ (labels near the curves), Vegard coefficient $W = -1$ Å$^3$ **(c)** and $W = +1$ Å$^3$ **(d)**. Potential and polarization profiles are almost indifferent to $W$ values in the range {-1,+1} Å$^3$. BFO polarization $P_S = +0.9$ C/m$^2$, $\alpha = -0.29\times10^9$ F/m, elastic compliances $s_{11}$=5.29 $10^{-12}$ Pa$^{-1}$, $s_{12}$= − 1.85 $10^{-12}$ Pa$^{-1}$, electrostriction $Q_{11}$=0.032, $Q_{12}$= − 0.016 C$^4$/m$^2$ [32], extrapolation length $\lambda_P = 0.1L_P$; $k_BT$ =0.025 eV (room), equilibrium concentration $N_d^0 = 10^{25}$ m$^{-3}$, typical values of the level difference $E_d - E_C = -0.1$ eV. Fermi level was defined in a self-consistent way [51, 52, 53, 54]

**Figures 3a** and **3b** demonstrate how the screening radius depends on the misfit strain $u_m$ and Vegard coefficient $W$ respectively. **Figure 3a** shows that the screening radius $R_d$ is independent on misfit strain $u_m$ in the case of zero Vegard coefficient, while non-zero values of the coefficient (both positive and negative) leads to the pronounced nonlinear dependence of $R_d$ on $u_m$, that is in qualitative agreement with Eqs.(11)-(12). One can see from **Figure 3b** that $R_d$ value nonlinearly depends on the Vegard coefficient and the curves calculated at different $u_m$ cross at a zero point $W$=0 (as anticipated from the **Figure 3a**). The dashed line in **Figure 3b** is calculated for $u_m = 0$; so it shows the bare impact of the electrostriction coupling and Vegard effect on the screening radius. Approximate analytics gives $R_d = R_d^0 \exp\left(\frac{W Q_{12} \overline{P_3^2}}{k_B T (s_{11} + s_{12})}\right)$ for the case $u_m = 0$. Solid curves demonstrate how strongly (up to 10 times) compressive misfit strains can influence on the screening radius dependence on Vegard coefficient.

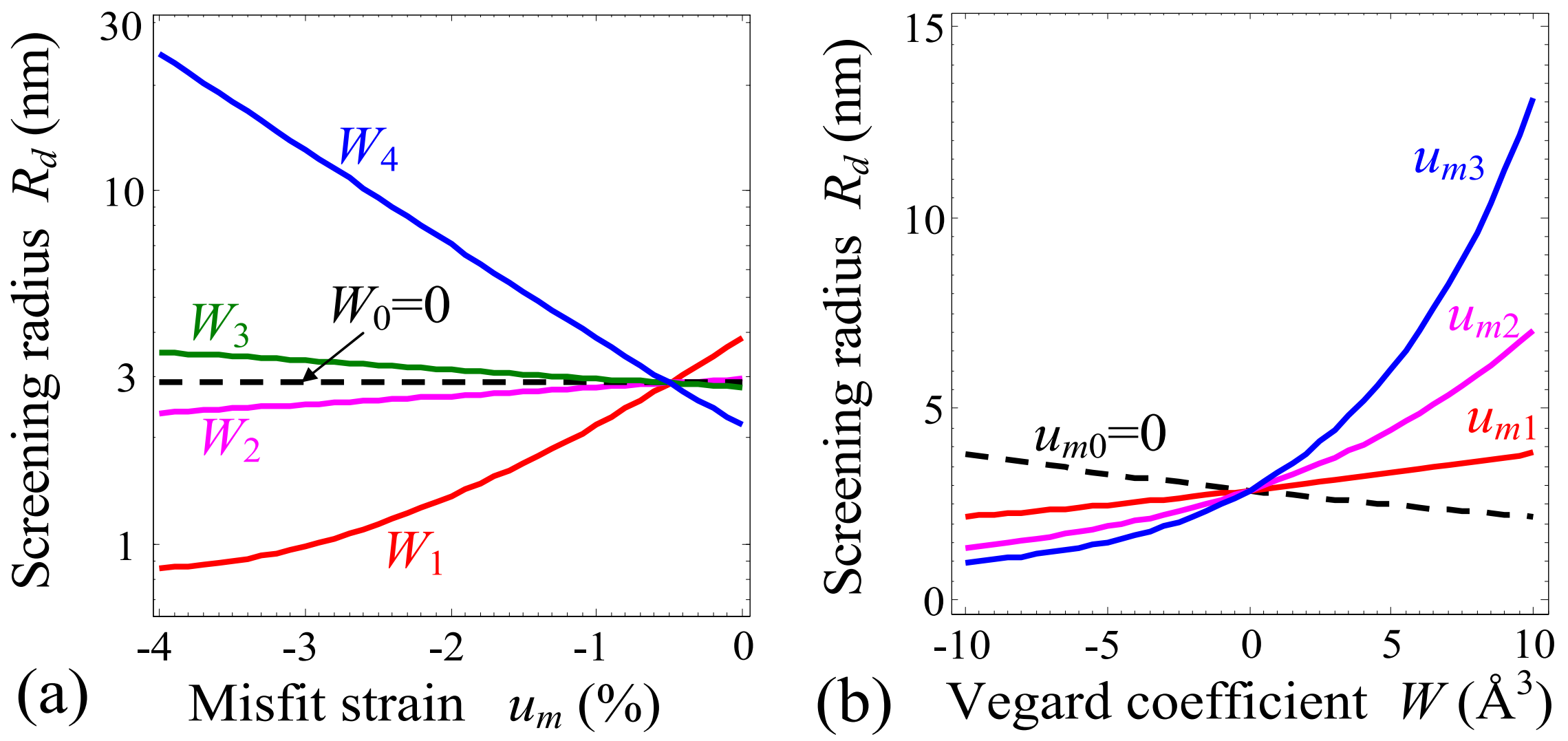

**Figure 3. (a)** Dependence of the screening radius $R_d$ on the misfit strain $u_m$ calculated for different Vegard coefficients $W_0 = 0$, $W_1 = -10$ Å$^3$, $W_2 = -1$ Å$^3$, $W_3 = +1$ Å$^3$, $W_4 = +10$ Å$^3$ (labels near the curves). **(b)** $R_d$ vs. Vegard coefficient $W$ calculated for different misfit strain $u_{m0} = 0$, $u_{m1} = -1\,\%$, $u_{m2} = -2\,\%$, $u_{m3} = -3\,\%$, (labels near the curves). BFO parameters are the same as in the **Figure 2**.

Now let us study the dependence of the donor concentration averaged over the film thickness ($\overline{N}_d^+$) and equilibrium concentration ($N_{de}^+$) on misfit strain and Vegard stress. Dependences of the relative concentrations $\overline{N}_d^+/N_d^0$ and $N_{de}^+/N_d^0$ on the Vegard coefficient $W$ for different misfit strains are shown in **Figures 4a-b**. The dependence of $\overline{N}_d^+/N_d^0$ on $W$ is linear for $u_m=0$ and quasi-linear with smooth oscillations for $u_m<0$. The concentrations $\overline{N}_d^+$ and $N_{de}^+$ increase with $|u_m|$ increase. The concentration $N_{de}^+$ is the highest for negative $W$ and decreases when the coefficient becomes positive. Importantly that depending on the sign of Vegard effect and misfit strain value equilibrium donor concentration changes dramatically [**Figure 4b**].

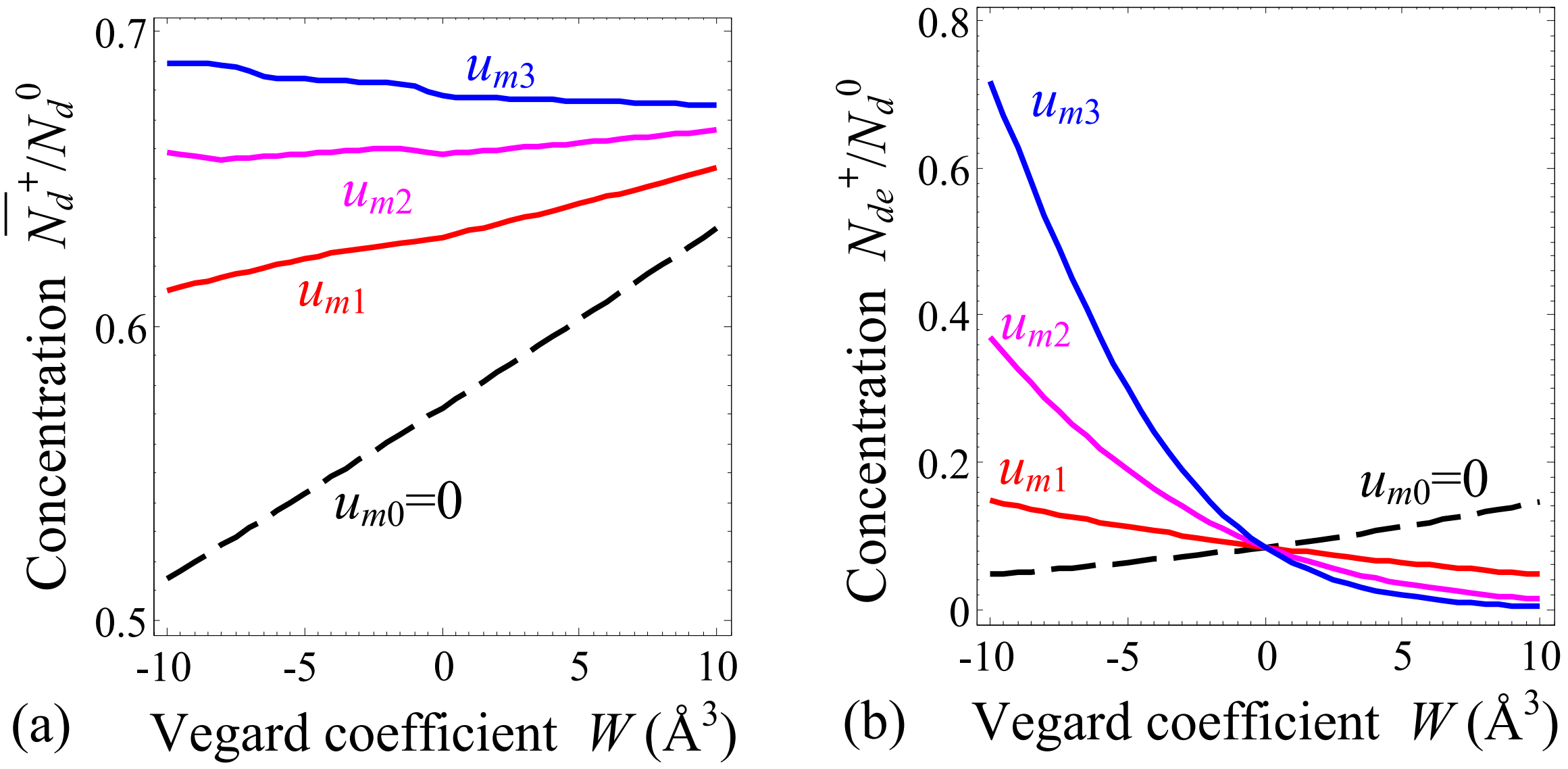

**Figure 4.** Dependences of the average and equilibrium concentrations of ionized donors, $\overline{N}_d^+/N_d^0$ **(a)** and $N_{de}^+/N_d^0$ **(b)**, on the Vegard coefficient W calculated for a set of misfit strains $u_{m0}=0$, $u_{m1}=-1\%$, $u_{m2}=-2\%$ and $u_{m3}=-3\%$ (labels near the curves). $BiFeO_3$ film thickness $h$= 20 nm, other parameters are the same as in **Figure 2.**

On further analysis, it appears that at least three physical mechanisms of the *misfit strain-driven defect gradient in thin strained ferroelectric films* are present. ***The first driving force*** is the inhomogeneous internal electric field originated from the depolarization effects related with spontaneous polarization abrupt at the film surfaces and from the built-in potential (if any originated at the film-substrate interface). The electric field potential gradient causes the charged defects concentration variation $N(x_3) \propto N^0 \exp\left(-\frac{eZ_d}{k_B T}\varphi(x_3)\right)$. The region of the defect gradient near the film-substrate interface can be estimated as the Debye screening radius $R_d$ that is exponentially dependent on misfit strain $R_d \propto \exp\left(-\frac{Wu_m}{2(s_{11}+s_{12})k_B T}\right)$. For $u_m = 0$ typical $R_d$ values are (1-10) nm at room temperature. The exponential dependence of Debye screening radius on $u_m$ changes the scale up to several orders of magnitude.[20] ***The second driving force*** is the intrinsic stress gradient $\partial\sigma_0/\partial x_3$ across the film-substrate interface. The stress gradient can be caused by the polarization gradient across the interface via electrostriction $Q_{ij}$ and flexoelectric effect $F_{ij}$: $\frac{\partial\sigma_0}{\partial x_3} \propto -\frac{2Q_{12}P_3}{s_{11}+s_{12}}\frac{\partial P_3}{\partial x_3} + \frac{2F_{13}}{s_{11}+s_{12}}\frac{\partial^2 P_3}{\partial x_3^2}$. The polarization component normal to the interface is suppressed by the depolarization field, and thus its gradient in ferroelectric material without free carriers can extend only for 0.1 – 0.5 nm, since the corresponding correlation length $L_p$ is small, $L_P \propto 0.1 - 0.2$ nm. However, in the case of ferroelectric-semiconductor the actual scale of polarization gradient is determined by the screening space charge layer width that can be much higher than $L_P$. The octahedral tilt gradient across the interface can also contribute into the stress gradient in an antiferrodistortive ferroelectric like BFO via rotostriction tensor $R_{ij}$ as $\frac{\partial\sigma_0}{\partial x_3} \propto -\frac{2R_{12}\Phi_3}{s_{11}+s_{12}}\frac{\partial\Phi_3}{\partial x_3}$. The tilts variation are not suppressed by any analogy of electric depolarization field and thus its gradient can extend several correlation lengths $L_\Phi \propto 2 - 4$ nm, that gives a value ~ 10 nm.

***The third driving force*** is the free energy gain in the parent phase. Let us consider the model situation when defects are not charged ($Z_d = 0$), spontaneous polarization and tilt are absent in the parent high temperature phase corresponding to the deposition conditions. The variation of the Helmholtz free energy density is $\delta F = \frac{\left(u_m - W\left(N(x_3) - N^0\right)\right)^2}{s_{11}+s_{12}}$ (see **Appendix C** of the **Suppl. Mat.**) Since always $(s_{11}+s_{12}) > 0$, the variation is positive. So defect redistribution becomes the most thermodynamically favorable when the variation $\delta N(x_3) = u_m/W$ is homogeneous. Neutral defects try to reach homogeneous distribution inside the film in paraelectric phase, but the equilibrium state may not be reached in a real time scale (e.g. cation diffusion) due to the associated long duration of the kinetics. Adding temporal kinetics makes the system behaviour more complex as will be demonstrated in the next section.

## IV. Ferroelectric polarization reversal and electric current kinetics

In dimensionless variables, the kinetic equations for donors and electrons, their chemical potentials (2)-(5) coupled with Poisson equation for electric potential (1), time-dependent LGD equation (8) and corresponding boundary conditions are listed in **Appendix B of Suppl. Mat.** The combined analysis reveals that the peculiarities of the system kinetics are defined by the Landau-Khalatnikov phonon time, $t_{LKh} = \Gamma/\alpha_T T_C^b$, characteristic electronic and donor times, $t_e = L_P^2/(e\eta_e k_B T)$ and $t_d = L_P^2/(e\eta_d k_B T)$ respectively, where $L_P = \sqrt{g/\alpha_T T_C^b}$ is the correlation length at zero temperature. For numerical calculations performed in the MathLab package it is convenient to introduce the dimensionless total electric current $\tilde{J} = J/J_0$, [55] polarization $\tilde{P} = P_3/P_S$, potential $\psi = eU/(k_B T)$, external electric field $\tilde{E} = \psi L_P/h$ and film thickness $\tilde{h} = h/L_P$. The current scale $J_0 = e N_{de}^+ \eta_e k_B T/L_P$, where $N_{de}^+ = N_d^0(1 - f((E_d - E_F)/k_B T))$ is the equilibrium concentration of ionized donors at zero potential and stress. Polarization scale is the spontaneous polarization at zero K, $P_S = \sqrt{\alpha_T T_C^b/\beta}$. External electric field is introduced as $E = U/h$ and its scale is the thermal activation-related field $E_T = k_B T/(eL_P)$.

Based on the above formulations, a numerical solution of the coupled system was carried out for the case of completely electron-conducting electrodes ($\xi \to \infty$ and $n_b = \overline{N}_d^+$), different characteristic times ratios $t_{LKh}/t_d$ and $t_e/t_d$, high enough compressive misfit strains $u_m = -1, -1.5, -2$ % and film thicknesses $\tilde{h} = 5 - 20$.

The ferroelectric polarization and electric current hysteresis loops calculated for different compressive misfit strains $u_m = -1, -1.5, -2$ %, film thickness $\tilde{h} = 5$ and hierarchy of characteristic times $t_{LKh} << t_e << t_d$ are shown in **Figure 5**. Note that the smallest thickness $\tilde{h} = 5$ was on purpose chosen to be very close to the critical thickness of the size-induced phase transition from the ferroelectric to the paraelectric state.[56, 57] In **Figure 5a**, the square-to-slim transition of the polarization hysteresis loop shape appears under misfit strain decrease. Also, it can be seen from **Figure 5a** and **5b** that the polarization hysteresis loops calculated for $u_m = -1\,\%$ are paraelectric. Polarization hysteresis loops calculated for $u_m = -1.5\,\%$ are non-linear paraelectric-like for slow varying external voltage with period $\tau_1 = 10t_d$, but become slim ferroelectric-like for the period $\tau_2 = t_d$, when the external voltage varies 10 times more rapidly.

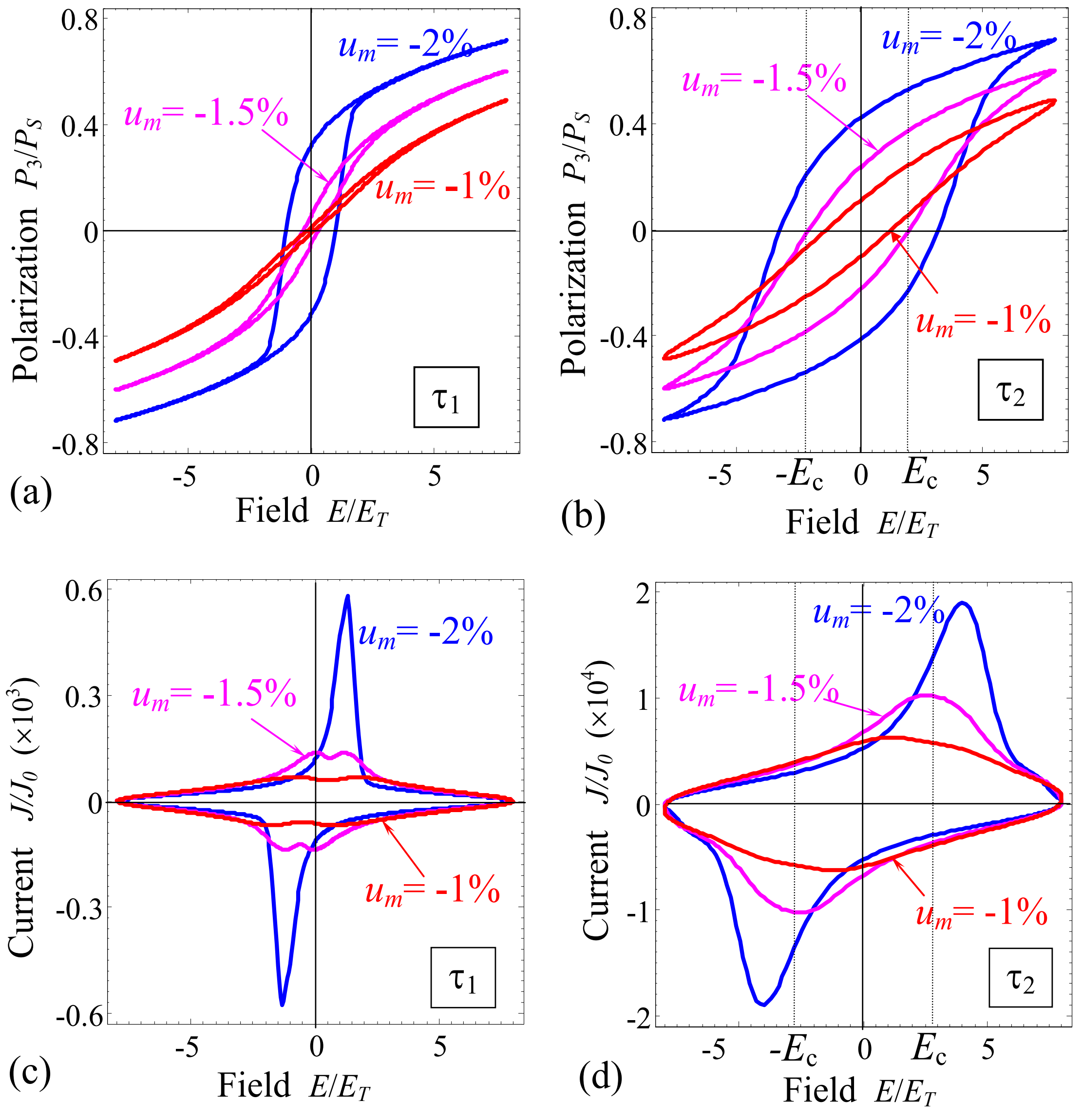

**Figure 5.** Dynamic polarization **(a,b)** and electric current **(c,d)** hysteresis loops calculated from Eqs.(9)-(12) for different applied voltage periods $\tau_1 = 10t_d$ **(a,c)** and $\tau_2 = t_d$ **(b,d)**, film thickness $\tilde{h} = 5$ and different misfit strains $u_m = -1, -1.5, -2\,\%$ (labels near the loops). Characteristic times $t_e = 0.1t_d$ and $t_{LKh} = 0.01t_d$. Vegard coefficients $W = -1\,Å^3$, extrapolation length $\lambda_P = 0.1L_P$. All other parameters are the same as in **Figure 2.**

Corresponding electric current curves with doubled smeared maxima at zero voltage confirm the paraelectric-like state of the film at $\tau_1 = 10t_d$ (**Figure 5c**), but for the case $\tau_2 = t_d$ the similar electric current curve shows a broader single maxima profile indicating the ferroelectric-like state of the same film (**Figure 5d**). Now this leads to the question, what happens with the loops of same film for the case of $\tau_2 = t_d$, while for the case $\tau_1 = 10t_d$ the film state is clearly non-linear paraelectric. In accordance with our model exactly the retarding of mobile donors with respect to applied voltage phase causes the dynamic hysteretic non-linearity of polarization change and, thus, induces ferroelectric-like behaviour. The behaviour is, in fact, illusive, because the coercive field for the polarization loop (defined as the loop intersection with *E*-axes) is smaller that for the current one. The coercive field for the current loop corresponds to the inflection point on the polarization loop. In other words, ferroelectric-like polarization behaviour for the film comes from the cation migration.

The ferroelectric polarization and electric current hysteresis loops calculated for fixed misfit strain $u_m = -1.5\%$, different film thickness $\tilde{h} = 5$, 10, 20 and hierarchy of characteristic times $t_{LKh} << t_e << t_d$ are shown in **Figure 6**. Qualitative similarity between the film behaviour under the misfit increase (**Figure 5**) and thickness increase (**Figure 6**) are evident. Namely, the square-to-slim transition of the polarization loop shape appears under the film thickness decrease from $\tilde{h} = 20$ to 5. It can be observed from **Figure 6a** and **6b** that the polarization hysteresis loops calculated for $\tilde{h} = 5$ are non-linear paraelectric-like for slow varying external voltage with period $\tau_1 = 10t_d$, but become slim ferroelectric-like for the period $\tau_2 = t_d$, when the external voltage varies 10 times more rapidly. Corresponding electric current curves with smeared maxima at zero voltage confirm the paraelectric-like state of the film at $\tau_1 = 10t_d$ (**Figure 6c**), but for the case $\tau_2 = t_d$ the coercive field $E_c$ appears indicating the ferroelectric-like state of the same film (**Figure 6d**). Here the ferroelectric-like polarization behaviour for the thinnest film comes from the donor migration. The film becomes a pronounced ferroelectric with the thickness increase, namely for thicknesses $\tilde{h} \geq 10$ we see a square-like ferroelectric hysteresis loop with pronounced coercive field and sharp maxima of electric current exactly at the coercive field. As anticipated from the LGD-description of the finite size effects in ferroelectric films [56, 57] the remanent polarization and coercive voltage increase with the film thickness increase (compare the loops for $\tilde{h} = 10$ and $\tilde{h} = 20$ in **Figure 6a,b**). The amplitude and sharpness of electric current maxima also increases with the film thickness increase (compare the loops for $\tilde{h}=10$ and $\tilde{h}=20$ in **Figure 6c,d**).

Complementary to **Figure 6, Figure 7** illustrates polarization and current loops for another hierarchy of characteristic times $t_{LKh}<t_e<t_d$ and different film thickness $\tilde{h}=5$, 10, 20. In comparison with **Figure 6** the same loops in **Figure 7** are changing in much more complicated way, but the film with thickness $\tilde{h}=5$ mimics ferroelectric-like behaviour even for slow varying voltage with a period $\tau_1=10t_d$.

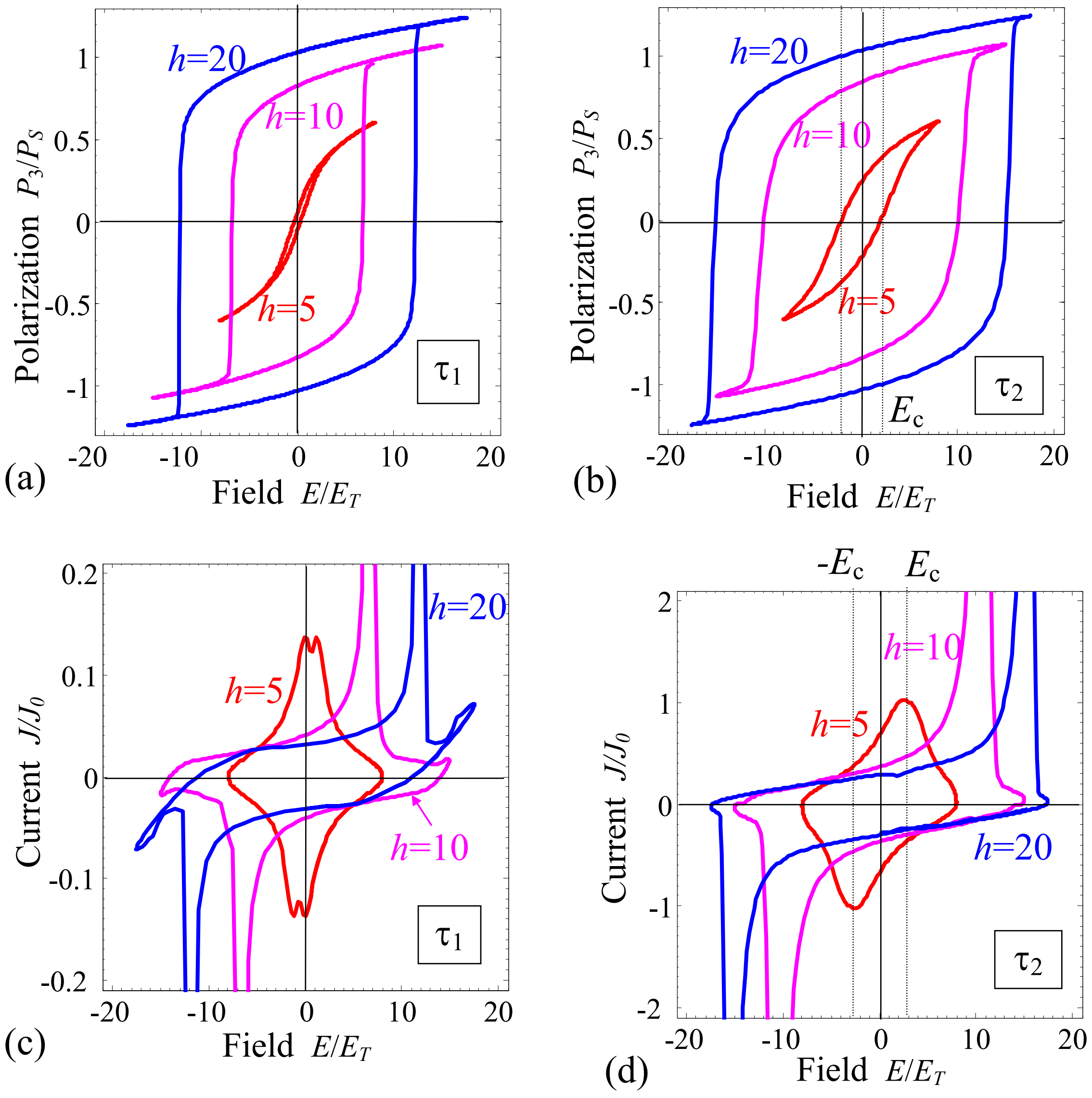

**Figure 6.** Dynamic polarization **(a,b)** and electric current **(c,d)** hysteresis loops calculated from Eqs.(9)-(12) for different applied voltage periods $\tau_1=10t_d$ **(a,c)** and $\tau_2=t_d$ **(b,d)**, film thickness $\tilde{h}=5$, 10, 20 (labels near the loops). Characteristic times $t_e=0.1t_d$ and $t_{LKh} = 0.01 t_d$. Misfit strain $u_m = -1.5\,\%$ and Vegard coefficients $W = -1\,\text{Å}^3$, extrapolation length $\lambda_P = 0.1 L_P$. All other parameters are the same as in **Figure 2.**

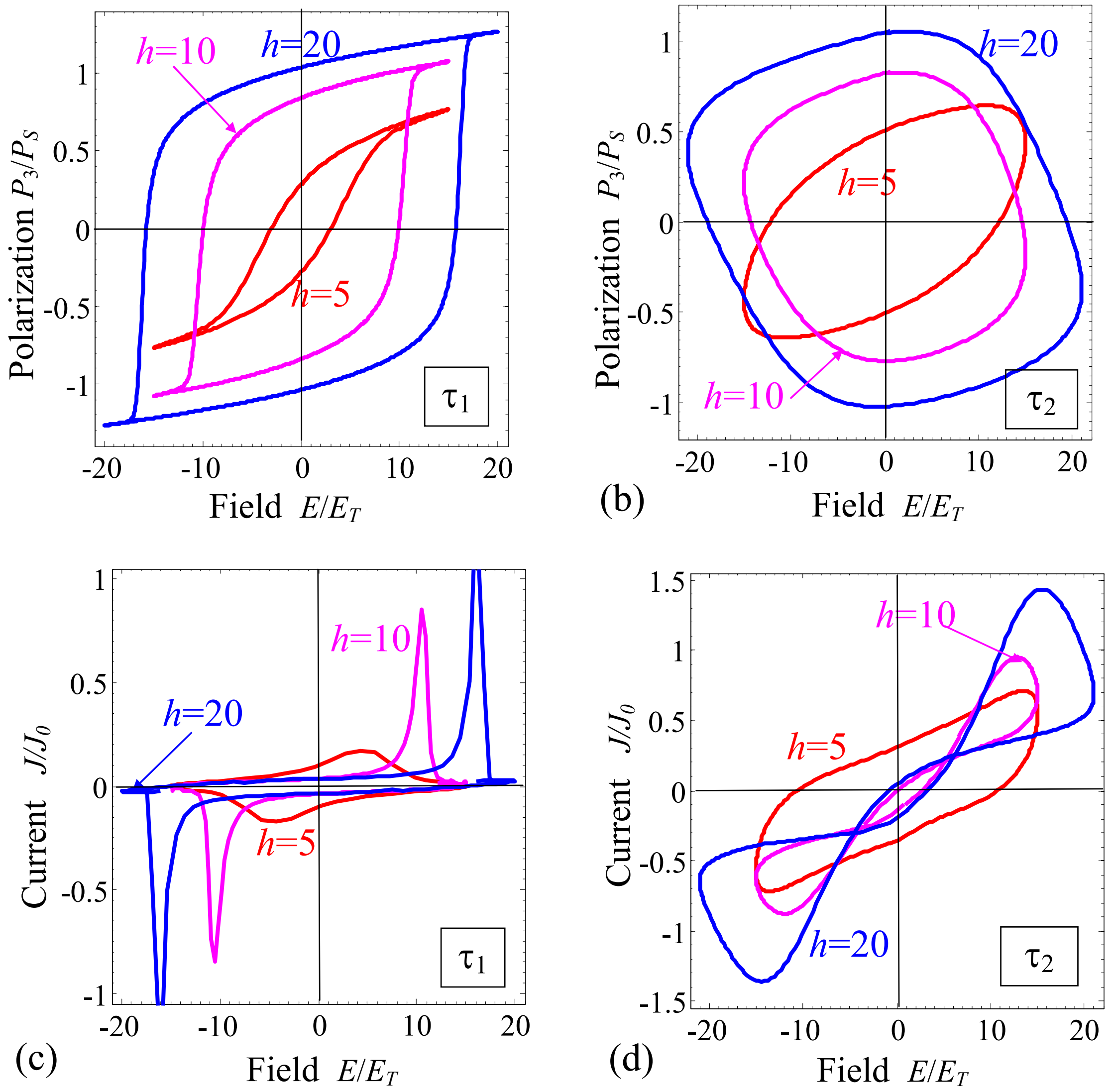

**Figure 7.** Dynamic polarization **(a,b)** and electric current **(c,d)** hysteresis loops calculated from Eqs.(9)-(12) for different applied voltage periods $\tau_1 = 10 t_d$ **(a,c)** and $\tau_2 = t_d$ **(b,d)**, film thickness $\tilde{h} = 5$, 10, 20 (labels near the loops). Characteristic times $t_e = 0.3 t_d$, $t_{LKh} = 0.1 t_d$. Misfit strain $u_m = -1.5\,\%$ and Vegard coefficients $W = -1\,\text{Å}^3$, extrapolation length $\lambda_P = 0.1 L_P$. All other parameters are the same as in **Figure 2.**

To summarize, the results of the section we underline that conventional Landau-Khalatnikov dynamics of ferroelectric polarization and electric current hysteresis loops takes place only at a hierarchy of characteristic times $t_{LKh} << t_e << t_d$ (**Figures 5, 6**), but it can be noticeably disturbed by the donor subsystem, once the inequality becomes not strict, e.g. at $t_{LKh} < t_e < t_d$ (**Figure 7**). Irrespective of the relationship between the time scales ($t_{LKh}$, $t_e$ and $t_d$) the changes in film thickness also have a strong influence on the hysteresis loop of electric current and polarization. But, the current loops undergo more pronounced changes than polarization loops under the film thickness increase. The shape of electric current hysteresis loops together with positions and absolute values of current maximum also changes in a significant way with the film thickness, misfit strain and the period of applied electric field. Hysteresis loops as a whole become narrower with the increasing of applied voltage period irrespective of the thickness and misfit strain. Further, the absolute value of the current maxima reduces with the decrease of film thickness irrespective of the given time scale. Eventually, it is worth noting that polarization and electric current loops shape calculated for the thinnest $BiFeO_3$ film are in a qualitative agreement with experimental results for ionic semiconductors.[58, 59] The observed square-to-slim transition of ferroelectric hysteresis with the decrease in misfit strain for fixed thickness or with the decrease in film thickness for fixed strain is in qualitative agreement with the reported theoretical and experimental results (see e.g. ref. [60]). Also, our modelling revealed that the system behaviour in kinetic regime is strongly dependent on the time scales and misfit strain for ultrathin films of thickness close to the critical thickness of the size-induced phase transition.

## V. Summary

We performed self-consistent modelling of the mobile donor kinetics and thermodynamics in thin ferroelectric films with misfit strain.

Calculations performed for the thermodynamic equilibrium state revealed that the driving forces of the defect segregation are different for charged and electro-neutral defects, namely:

1) The internal electric field (built-in and/or depolarization one) localized near the film-substrate interface leads to the strong accumulation or depletion of charged defects in the vicinity of the interface, at that corresponding screening length exponentially depends on misfit strain. Although the analytical results were derived in Boltzmann-Debye approximation, the main conclusions about the strong dependence of the physical quantities on misfit strain remained valid for the more general case of Fermi-Dirac statistics considered numerically.

2) The Vegard mechanism leads to the redistribution of both charge and electro-neutral defects in order to decrease effective stress in the strained films. Thus, the misfit strain can be the stimulus of the strong gradient of defect concentration in thin ferroelectric films clamped by substrate.
3) Importantly, our results can explain recent experimental observations [20]. Namely, for the case of BFO thin film epitaxially grown on LAO substrate our modelling predicts, that donor concentration decreases in 10 times at distances about 5-10 nm from the BFO/LAO interface.

Kinetics of polarization reversal and electric current are defined by the complex interplay between the donor, electron and phonon relaxation times complicated by the finite size effect. Namely:
1) Conventional Landau-Khalatnikov dynamics of ferroelectric polarization and electric current hysteresis loops takes place only when the phonon relaxation time is much smaller than the electron relaxation time, and, the electron relaxation is much smaller than the donor relaxation time. However, the Landau-Khalatnikov dynamics can be noticeably disturbed by the donor subsystem, when any of the time scale conditions are violated.
2) Hysteresis loops of electric current undergo more pronounced changes under the film thickness increase than polarization loops. The shape of electric current hysteresis loops together with positions and absolute values of current maximum also changes in a significant way with both the film thickness and the frequency of applied field.
3) The system behaviour in kinetic regime strongly depends on the time scales and misfit strain for ultrathin films.

Overall, a comprehensive approach on the behaviour of ferroelectric thin films under various influences such as misfit strain, chemical pressure, cation and donor migration and characteristic time scale of phonon relaxation were studied in detail and their critical influence on the functional behaviour were analysed. These results in tandem with the previously reported experimental results can be expected to provide more insight in-to the interfacial phenomena governing the functional oxide hetero-epitaxial thin films.

**Acknowledgements**
A.N.M. and E.A.E. acknowledge the support via bilateral SFFR-NSF project (US National Science Foundation under NSF-DMR-1210588 and State Fund of Fundamental State Fund of Fundamental Research of Ukraine, grant UU48/002). The research at UNSW was supported by an ARC Discovery grant. S.V.K. and A.Y.B. acknowledges Office of Basic Energy Sciences, U.S. Department of Energy.

**Supplementary Materials to**

# "Defect thermodynamics and kinetics in thin strained ferroelectric films: the interplay of possible mechanisms"

Anna N. Morozovska[1,2], Eugene A. Eliseev[2], P.S.Sankara Rama Krishnan[3], Alexander Tselev[4], Evgheny Strelkov[4], Albina Borisevich[4], Olexander V. Varenyk[5], Nicola V. Morozovsky[1], Paul Munroe[3], Sergei V. Kalinin[4*] and Valanoor Nagarajan[3†]

[1]Institute of Physics, [2]Institute of Problems for Material Sciences, NAS of Ukraine, 03028 Kiev, Ukraine

[3]School of Materials Science and Engineering, the University of New South Wales, Sydney, NSW2052 Australia.

[4]The Center for Nanophase Materials Sciences, Oak Ridge National Laboratory, Oak Ridge, TN 37922

[5]Taras Shevchenko Kiev National University, Radiophysical Faculty, 4g, pr. Akademika Hlushkova, 03022 Kiev, Ukraine

[*] Corresponding author

[†] Corresponding author

## Appendix A. LGD thermodynamic approach

Let us calculate the defects distribution behaviour in a ferroelectric film of thickness $h$ clamped by a thick rigid substrate. Axis $x_3$ is perpendicular to the film surface. $u_m$ is the film-substrate misfit strain. Gibbs potential bulk density is [1, 2, 3]:

$$G = G_{pol} + G_{tilt} + G_{elastic-striction} + G_{flexo} + G_{electric-vac}, \tag{A.1a}$$

$$G_{pol} = a_i P_i^2 + a_{ij} P_i^2 P_j^2 + a_{ijk} P_i^2 P_j^2 P_k^2 + \frac{g_{ijkl}}{2}\frac{\partial P_i}{\partial x_j}\frac{\partial P_k}{\partial x_l} - P_i E_i, \tag{A.1b}$$

$$G_{str} = b_i \Phi_i^2 + b_{ij}\Phi_i^2\Phi_j^2 + b_{ijk}\Phi_i^2\Phi_j^2\Phi_k^2 + \frac{v_{ijkl}}{2}\frac{\partial \Phi_i}{\partial x_j}\frac{\partial \Phi_k}{\partial x_l}, \tag{A.1c}$$

$$G_{elastic-striction} = -Q_{ijkl}\sigma_{ij}P_k P_l - R_{ijkl}\sigma_{ij}\Phi_k\Phi_l - \frac{s_{ijkl}}{2}\sigma_{ij}\sigma_{kl}, \tag{A.1d}$$

$$G_{flexo} = \frac{F_{ijkl}}{2}\left(\frac{\partial P_k}{\partial x_l}\sigma_{ij} - P_k\frac{\partial \sigma_{ij}}{\partial x_l}\right) \tag{A.1e}$$

$$G_{electric-vac} = -\sum_k \left(Z_k e\varphi N_k - W_{ij}^k\left(N_k - \langle N_k\rangle\right)\sigma_{ij} + k_B T\, S\left(N_k^0, N_k\right)\right) - \frac{f_k}{2}\left(\frac{\partial N_k}{\partial x_i}\right)^2. \tag{A.1f}$$

Here $P_i$ are polarization components, $\Phi_i$ is the structural order parameter (oxygen octahedral tilt), elastic stress tensor is $\sigma_{ij}$, $Q_{ijkl}$ are electrostriction tensor coefficients, $R_{ijkl}$ are the rotostriction tensor coefficients, $F_{ijkl}$ is the flexoelectric coupling tensor, $g_{ijkl}$ and $v_{ijkl}$ are gradient coefficients tensor, $\varphi$ is the electric potential, electric field is $E_i = -\partial\varphi/\partial x_i$. $e$=1.6×10$^{-19}$ C the electron charge, $Z_k$ is the defect charge (that is equal to zero for uncharged vacancies or isovalent impurities). Mobile charged species (e.g. substitution and/or interstitial defects, vacancies) equilibrium concentration is $\langle N_k\rangle$, its variation is $\delta N_k = N_k - \langle N_k\rangle$, defects equilibrium concentration is $\langle N_k\rangle << 2.25\times10^{28}$ m$^{-3}$. Vegard strain tensor $W_{ij}^k = W^k\delta_{ij}$ [4, 5, 6] is regarded diagonal ($\delta_{ij}$ is delta Kroneker symbol). In a general case the structure of Vegard expansion tensor (elastic dipole) is controlled by the symmetry (crystalline or Curie group symmetry) of the material. Configurational entropy in Boltzmann approximation is $S(x,y) = y\ln(y/x) - y$.

Euler-Lagrange equation for ferroelectric polarization and tilts can be obtained by the Gibbs potential variation on polarization components:

$$2a_i P_i - Q_{klij}\sigma_{kl}P_j + 2a_{ij}P_i P_j^2 + 2a_{ijk}P_i P_j^2 P_k^2 - g_{ijkl}\frac{\partial^2 P_k}{\partial x_j \partial x_l} = E_i \tag{A.2a}$$

$$2b_i\Phi_i - R_{klij}\sigma_{kl}\Phi_j + 2b_{ij}\Phi_i\Phi_j^2 + 2b_{ijk}\Phi_i\Phi_j^2\Phi_k^2 - v_{ijkl}\frac{\partial^2 \Phi_k}{\partial x_j \partial x_l} = 0 \tag{A.2b}$$

The potential φ should be determined self-consistently from the Poisson equation:

$$\varepsilon_0\varepsilon_b\frac{\partial^2\varphi}{\partial x_i^2} = \frac{\partial P_i}{\partial x_i} - e\left(\sum_k Z_k N_k - n\right) \tag{A.3}$$

$\varepsilon_0=8.85\times10^{-12}$ F/m the dielectric permittivity of vacuum; $\varepsilon_b$ is a background permittivity,[7] $n$ is the concentration of the electrons in the conduction band.

Further let us consider out-of-plane components of the tilt and polarization for the sake of simplicity. The answer for more general case is listed in the main text. Equations of state $\partial G/\partial\sigma_{ij} = -u_{ij}$ give that in-plane and out-of-plane strains $u_{ij}$:

$$u_{11} = s_{11}\sigma_{11} + s_{12}\sigma_{22} + s_{12}\sigma_{33} + Q_{12}P_3^2 + R_{12}\Phi_3^2 + W^k\,\delta N_k, \tag{A.4a}$$

$$u_{22} = s_{11}\sigma_{22} + s_{12}\sigma_{11} + s_{12}\sigma_{33} + Q_{12}P_3^2 + R_{12}\Phi_3^2 + W^k\,\delta N_k, \tag{A.4b}$$

$$u_{33} = s_{11}\sigma_{33} + s_{12}(\sigma_{22} + \sigma_{11}) + Q_{11}P_3^2 + R_{11}\Phi_3^2 + W^k\,\delta N_k, \tag{A.4c}$$

$$u_{12} = s_{44}\sigma_{12}, \quad u_{13} = s_{44}\sigma_{13}, \quad u_{23} = s_{44}\sigma_{23}. \tag{A.4d}$$

$s_{ij}$ are elastic compliances. For an epitaxial thin film clamped by a rigid substrate mechanical boundary conditions at $x_3 = 0$ and $x_3 = h$ are conventional:

$$u_{11}(h) = u_{22}(h) = u_m, \qquad \sigma_{3i}(0) = 0. \tag{A.5}$$

$u_m$ is the film-substrate misfit strain. For the one-dimensional distributions mechanical equilibrium equation is $\partial\sigma_{i3}/\partial x_3 = 0$. Next one could recall the conditions of elastic compatibility $\mathrm{inc}(i,j,\hat{u}) = e_{ikl}e_{jmn}u_{ln,km} = 0$. For the considered case they could be reduced to $\partial^2 u_{11}/\partial x_3^2=0$ (at that the distribution of $u_{33}$ can be arbitrary along $x_3$). Approximate solution is

$$u_{12} = u_{13} = u_{23} = 0 \qquad u_{11} = u_{22} = u_m \tag{A.6a}$$

$$u_{33} = Q_{11}P_3^2 + R_{11}\Phi_3^2 + W^k\,\delta N_k + \frac{2s_{12}\left(u_m - Q_{12}P_3^2 - R_{12}\Phi_3^2 - W^k\,\delta N_k\right)}{s_{11} + s_{12}} \tag{A.6b}$$

$$\sigma_{12} = \sigma_{13} = \sigma_{23} = \sigma_{33} = 0 \tag{A.6c}$$

$$\sigma_{11}=\sigma_{22}=\frac{u_m-W^k\,\delta N_k-Q_{12}P_3^2-R_{12}\Phi_3^2}{s_{11}+s_{12}} \tag{A.6d}$$

In the simplest case the stress is:

$$\sigma\equiv\sigma_{11}+\sigma_{22}=\sigma_0-\frac{2W^k\,\delta N_k}{s_{11}+s_{12}}. \tag{A.7a}$$

$$\sigma_0(x_3)=\frac{2u_m-(Q_{11kl}+Q_{22kl})P_k\,P_l}{s_{11}+s_{12}}+\frac{2F_{11k3}}{s_{11}+s_{12}}\frac{\partial P_k}{\partial x_3} \tag{A.7b}$$

Here $Q_{ijkl}$ is the electrostriction tensor coefficient, $P_i$ are the polarization components, $s_{ij}$ are elastic compliances. Flexoelectric coupling tensor is $F_{ijkl}$. Note that one can ignore the impact of the flexoelectric coupling on the shear stress only for the case of purely out-of-plane polarization $\mathbf{P}\equiv P_3(x_3)$.

## Appendix B. Ferroelectric polarization reversal and electric current kinetics in dimensionless variables

Using dependencies of concentration of donors, $N_d^+=N_d^0 f(-E_d-W\sigma+eZ_d\varphi-\zeta_d)$, and electrons, $n=N_C F_{1/2}((e\varphi+\zeta_e-E_C)/k_BT)$, on the electrochemical potentials $\zeta_{d,e}$, one can rewrite them via chemical potentials $\mu_d=eZ_d\varphi-\zeta_d-W\sigma$ and $\mu_e=e\varphi+\zeta_e$, as $N_d^+=N_d^0 f(\mu_d-E_d)$ and $n=N_C F_{1/2}((\mu_e-E_C)/k_BT)$. Then coupled equations (2) and (4) become $\frac{\partial N_d^+}{\partial t}-\frac{\partial}{\partial x_3}\left(\eta_d N_d^+\frac{\partial(eZ_d\varphi-\mu_d-W\sigma)}{\partial x_3}\right)=0$ and $\frac{\partial n}{\partial t}-\frac{\partial}{\partial x_3}\left(\eta_e n\frac{\partial(\mu_e-e\varphi)}{\partial x_3}\right)=0$.

In dimensionless variables the coupled equations for chemical potentials can be written as

$$\frac{t_d}{t_e}\frac{\partial\left(f(\tilde{\mu}_d-\tilde{E}_d)\right)}{\partial\tilde{t}}-\frac{\partial}{\partial\tilde{z}}\left(f(\tilde{\mu}_d-\tilde{E}_d)\frac{\partial}{\partial\tilde{z}}\left(\tilde{\varphi}-\tilde{\mu}_d-\left(\tilde{W}_m-\tilde{w}^2 f(\tilde{\mu}_d-\tilde{E}_d)\right)\right)\right)=0, \tag{B.1a}$$

$$\frac{\partial}{\partial\tilde{t}}\left(F_{1/2}(\tilde{\mu}_e-\tilde{E}_C)\right)-\frac{\partial}{\partial\tilde{z}}\left(F_{1/2}(\tilde{\mu}_e-\tilde{E}_C)\frac{\partial}{\partial\tilde{z}}(\tilde{\mu}_e-\tilde{\varphi})\right)=0. \tag{B.1b}$$

In the most compact form:

$$\frac{\partial f(\mu_d-E_d)}{\partial t}-\frac{\partial}{\partial x_3}\left(\eta_d f(\mu_d-E_d)\frac{\partial(eZ_d\varphi-\mu_d-W\sigma)}{\partial x_3}\right)=0 \tag{B.1c}$$

$$\frac{\partial}{\partial t}F_{1/2}\left(\frac{\mu_e-E_C}{k_BT}\right)-\frac{\partial}{\partial x_3}\left(\eta_e F_{1/2}\left(\frac{\mu_e-E_C}{k_BT}\right)\frac{\partial(\mu_e-e\varphi)}{\partial x_3}\right)=0. \tag{B.1d}$$

Dimensionless variables and parameters involved in Eqs.(B.1) are listed in the **Table 1.** From the table one could get the donor concentration and electron density as

$\tilde{N}=\left(N_d^+/N_{de}^+\right)=\left(N_d^0/N_{de}^+\right)f\left(\tilde{\mu}_d-\tilde{E}_d\right)$ and $\tilde{n}=\left(N_C/N_{de}^+\right)F_{1/2}\left(\tilde{\mu}_e-\tilde{E}_C\right)$, where the equilibrium concentration of ionized donors is $N_{de}^+=N_d^0\left(1-f\left(\left(E_d-E_F\right)/k_BT\right)\right)$, $N_d^0$ is the maximal steric concentration of defect atoms. Approximation for inverse Fermi integral is $F_{1/2}^{-1}(\tilde{n})\approx\left(3\sqrt{\pi}\tilde{n}/4\right)^{2/3}+\ln\left(\tilde{n}/(1+\tilde{n})\right)$.

Boundary conditions for currents in dimensionless variables are regarded donor-blocking and electron-conducting at least partially:

$$\left(\tilde{J}_d\right)\Big|_{\tilde{z}=0}=0;\quad \left(\tilde{J}_d\right)\Big|_{\tilde{z}=\tilde{h}}=0, \tag{B.2a}$$

$$\left(\tilde{J}_e-\tilde{\xi}\left(\tilde{n}-\tilde{n}_b\right)\right)\Big|_{\tilde{z}=0}=0 \qquad \left(\tilde{J}_e+\tilde{\xi}\left(\tilde{n}-\tilde{n}_b\right)\right)\Big|_{\tilde{z}=\tilde{h}}=0 \tag{B.2b}$$

Poisson equation for electric potential

$$\frac{\partial^2\tilde{\varphi}}{\partial\tilde{z}^2}=\frac{1}{\chi}\frac{\partial\tilde{P}}{\partial\tilde{z}}-\frac{L_P^2}{R_d^2}\left(\tilde{N}-\tilde{n}\right). \tag{B.3a}$$

Boundary conditions for electric potential

$$\left(\tilde{\varphi}\right)\Big|_{\tilde{z}=0}=\tilde{V}\sin\left(\tilde{\omega}\tilde{t}\right),\quad \left(\tilde{\varphi}\right)\Big|_{\tilde{z}=\tilde{h}}=0. \tag{B.3b}$$

Dynamic Euler-Lagrange equation for polarization belongs to the Landau-Kahalatnikov type. It has the form:

$$\frac{t_{LKh}}{t_e}\frac{\partial\tilde{P}}{\partial\tilde{t}}-\left(1-\frac{T}{T_C^b}+q_m+q_W\left(\tilde{N}-1\right)\right)\tilde{P}+\tilde{P}^3-\frac{\partial^2\tilde{P}}{\partial\tilde{z}^2}+\theta\frac{\partial\tilde{\varphi}}{\partial\tilde{z}}=0, \tag{B.4a}$$

Boundary conditions for polarization are:

$$\left(\tilde{P}-\frac{\lambda_P}{L_P}\frac{\partial}{\partial\tilde{z}}\tilde{P}\right)\Bigg|_{z=0}=0,\ \left(\tilde{P}+\frac{\lambda_P}{L_P}\frac{\partial}{\partial\tilde{z}}\tilde{P}\right)\Bigg|_{z=h}=0. \tag{B.4b}$$

All dimensionless variables and parameters involved in Eqs.(B.1)-(B.4) are also listed in the **Table S1.**

**Table S1. Dimensionless variables and parameters**

| **Quantity** | **Definition/designation** |
|---|---|
| Dimensionless coordinate and thickness | $\tilde{z}=x_3/L_P,\quad \tilde{h}=h/L_P$ |
| Correlation length at zero temperature | $L_P=\sqrt{g/\alpha_T T_C^b}$ |
| Curie temperature of a bulk material | $T_C^b$ |
| Dimensionless time | $\tilde{t}=t/t_e$ |
| Characteristic electron and donor times | $t_e=L_P^2/\eta_e k_B T$, $t_d=L_P^2/\eta_d k_B T$ |
| Dimensionless donor concentration | $\tilde{N}=N_d^+/N_{de}^+=\left(N_d^0/N_{de}^+\right)f\left(\tilde{\mu}_d-\tilde{E}_d\right)$ |
| Dimensionless electron density | $\tilde{n}=n/N_{de}^+=\left(N_C/N_{de}^+\right)F_{1/2}\left(\tilde{\mu}_e-\tilde{E}_C\right)$ |
| Equilibrium concentration of ionized donors at zero potential and stress | $N_{de}^+=N_d^0\left(1-f\left(\left(E_d-E_F\right)/k_BT\right)\right)$ |
| Dimensionless electric potential | $\tilde{\varphi} = e\varphi/k_B T$ |
| Applied voltage | $\tilde{V} = eU/k_B T$ |
| Dimensionless chemical potentials | $\tilde{\mu}_d = \mu_d/k_B T$ , $\tilde{\mu}_e = \mu_e/k_B T$ |
| Dimensionless donor level | $\tilde{E}_d \approx E_d/k_B T$ |
| Dimensionless conduction band position | $\tilde{E}_C = E_C/k_B T$ |
| Dimensionless Vegard coefficient coupled with misfit | $\tilde{W}_m = \dfrac{2u_m W}{(s_{11}+s_{12})k_B T}$, $\tilde{w}^2 = \dfrac{2W^2 N_d^0}{(s_{11}+s_{12})k_B T}$ |
| Dimensionless electron and donor currents | $\tilde{J}_e = \dfrac{L_P J_e}{N_{de}^+ \eta_e k_B T}$, $\tilde{J}_d = \dfrac{L_P J_d}{N_{de}^+ \eta_e k_B T}$ |
| Dimensionless rate constant and concentration | $\tilde{\xi} = \dfrac{L_P}{\eta_e k_B T}\xi$, $\tilde{n}_b = \dfrac{n_b}{N_{de}^+}$ |
| Debye screening radius | $R_d = \sqrt{\varepsilon_0 \varepsilon_{33}^b k_B T/(e^2 N_{de}^+)}$ |
| Characteristic temperatures | $\chi = \dfrac{\varepsilon_0 \varepsilon_{33}^b k_B T}{e L_P P_S}$, $\theta = \dfrac{k_B T}{\alpha_T T_C e P_S L_P}$ |
| Landau-Khalatnikov time | $t_{LKh} = \Gamma/\alpha_T T_C^b$ |
| Dimensionless polarization | $\tilde{P} = P_3/P_S$ |
| Spontaneous polarization at zero K | $P_S = \sqrt{\alpha_T T_C^b/\beta}$ |
| Dimensionless electrostrictive coupling | $q_m = \dfrac{4Q_{12}u_m}{T_C^b \alpha_T (s_{11}+s_{12})}$, $q_W = \dfrac{4Q_{12}W N_{de}^+}{T_C^b \alpha_T (s_{11}+s_{12})}$ |
| Spontaneous polarization at temperature $T$ | $P_S^T = \sqrt{\alpha_T (T_C^b - T)/\beta}$ |
| Thermodynamic coercive field | $E_{cr}^T = (2/3\sqrt{3})\alpha_T (T_C^b - T) P_S^T$ |

Numerical solution of the system (9)-(12) was done for the case of completely electron-conducting electrodes with ($\tilde{\xi} \to \infty$, $\tilde{n}_b = 1$), different film thickness $\tilde{h} = 5 - 20$ and characteristic times ratios $t_{LKh}/t_d$ and $t_e/t_d$ .

Approximation for inverse Fermi integral is $F_{1/2}^{-1}(\tilde{n}) \approx \left(3\sqrt{\pi}\tilde{n}/4\right)^{2/3} + \ln(\tilde{n}/(1+\tilde{n}))$ was derived from the expansion

$$F_{1/2}(\xi) = \frac{2}{\sqrt{\pi}} \int_0^\infty \frac{\sqrt{\zeta}}{1+\exp(\zeta - \xi)} d\zeta \approx \begin{cases} \exp(\xi), & \xi \to -\infty \\ \dfrac{4}{3\sqrt{\pi}}\xi^{3/2} \end{cases} \propto \left( \exp(-\xi) + \frac{3\sqrt{\pi}}{4}\left(4+\xi^2\right)^{-3/4} \right)^{-1} \qquad \text{(B.5)}$$

**Appendix C. Equilibrium concentration of neutral defects in the parent phase**

Finally let us consider the model situation when defects are not charged ($Z = 0$), spontaneous polarization and tilt are absent in the parent high temperature phase corresponding to the deposition conditions. Corresponding part of the energy bulk density is:

$$G = -\frac{s_{ijkl}}{2}\sigma_{ij}\sigma_{kl} - W_{ij}\left(N(z) - \langle N \rangle\right)\sigma_{ij} \tag{C.1}$$

Equations of state $\partial G/\partial \sigma_{ij} = -u_{ij}$ give that in-plane and out-of-plane strains are:

$$u_{11}(z) = s_{11}\sigma_{11} + s_{12}\sigma_{22} + s_{12}\sigma_{33} + W\,\delta N(z), \tag{C.2a}$$

$$u_{22}(z) = s_{11}\sigma_{22} + s_{12}\sigma_{11} + s_{12}\sigma_{33} + W\,\delta N(z), \tag{C.2b}$$

$$u_{33}(z) = s_{11}\sigma_{33} + s_{12}(\sigma_{22} + \sigma_{11}) + W\,\delta N(z). \tag{C.2c}$$

Here all designations are the same as in **Appendix A**. Boundary conditions for strains in a thick single-domain film are $u_{11}(0) = u_{22}(0) = u_m$, $u_m$ is misfit strain, and $\sigma_{3i}(h) = 0$. Keeping in mind that $u_{11}(z) = u_{22}(z)$, solutions for the stresses derived from Eq.(C.2) are:

$$\sigma_{11} = \sigma_{22}(z) = (c_{11} + c_{12})(u_{11}(z) - W\,\delta N_m(z)) + c_{12}(u_{33}(z) - W\,\delta N_m(z)), \tag{C.3a}$$

$$\sigma_{33}(z) = 2c_{12}(u_{11}(z) - W\,\delta N_m(z)) + c_{11}(u_{33}(z) - W\,\delta N_m(z)), \tag{C.3b}$$

Where $c_{11} = \dfrac{s_{11} + s_{12}}{(s_{11} - s_{12})(s_{11} + 2s_{12})}$, $c_{12} = \dfrac{-s_{12}}{(s_{11} - s_{12})(s_{11} + 2s_{12})}$.

Mechanical equilibrium condition $\partial \sigma_{ij}/\partial x_j = 0$ gives Lame equation:

$$\partial \sigma_{33}/\partial z = 0. \tag{C.4}$$

Equation (5) leads to $\sigma_{33} = const$ that allowing for the boundary conditions $\sigma_{3i}(h) = 0$ gives

$$\sigma_{33} = 0. \tag{C.5}$$

Mechanical compatibility conditions in 1D-case gives for the film infinite in x,y directions:

$$u_{11}(z) = u_{22}(z) = u_m \tag{C.6}$$

Using Eqs.(C.5)-(C.6), Eqs.(C.3) acquire the form of equations for $\sigma_{11}(z) = \sigma_{22}(z)$ and $u_{33}(z)$:

$$0 = 2c_{12}(u_m - W\,\delta N_m(z)) + c_{11}(u_{33}(z) - W\,\delta N_m(z)), \tag{C.7a}$$

$$\sigma_{11}(z) = \sigma_{22}(z) = (c_{11} + c_{12})(u_m - W\,\delta N_m(z)) + c_{12}(u_{33}(z) - W\,\delta N_m(z)), \tag{C.7b}$$

From Eqs.(C.7) nonzero stresses and out-of-plane strain are

$$\sigma_{22}(z) = \sigma_{11}(z) = \frac{u_m - W\delta N_m(z)}{s_{11} + s_{12}}, \tag{C.8a}$$

$$u_{33}(z) = \frac{2s_{12}u_m}{s_{11}+s_{12}} + \left(\frac{s_{11}-s_{12}}{s_{11}+s_{12}}\right) W\delta N_m(z) \qquad \text{(C.8b)}$$

Substituting stresses (C.5), (C.8a), strains (C.6), (C.8b) and $\sigma_{12} = 0$ in the free energy (C.1) after Legendre transformation, $F = G + u_{11}\sigma_{11} + u_{22}\sigma_{22} + u_{12}\sigma_{12}$, after elementary calculations we obtain the variation of the free energy density:

$$\delta F = \begin{cases} \dfrac{(u_m - W\delta N(z))^2}{s_{11}+s_{12}}, & \textit{defects exist} \\ \dfrac{(u_m)^2}{s_{11}+s_{12}}, & \textit{no defects} \end{cases} \qquad \text{(C.9)}$$

Since always $(s_{11}+s_{12}) > 0$, the variation is positive and the ratio

$$\frac{\delta F(N \neq 0)}{\delta F(N = 0)} = \left(1 - \frac{W\delta N(z)}{u_m}\right)^2. \qquad \text{(C.10)}$$

So defects redistribution becomes the most thermodynamically favorable when $u_m \equiv W\delta N(z)$, i.e. when $\delta N(z) = \frac{u_m}{W}$ is homogeneous, but not gradient. On the other hand one can conclude from Eq.(C.10) neutral defects try to reach homogeneous distribution inside the film in paraelectric phase, but the equilibrium may not be reached in a real time scale due to high times of e.g. cation diffusion.